# Ultrafast depinning of domain wall in notched antiferromagnetic nanostructures

Z. Y. Chen[1], M. H. Qin[1,*], and J. –M. Liu[2]

[1]*Institute for Advanced Materials, South China Academy of Advanced Optoelectronics and Guangdong Provincial Key Laboratory of Quantum Engineering and Quantum Materials, South China Normal University, Guangzhou 510006, China*

[2]*Laboratory of Solid State Microstructures and Innovative Center for Advanced Microstructures, Nanjing University, Nanjing 210093, China*

[**Abstract**] The pinning/depinning of antiferromagnetic (AFM) domain wall is certainly the core issue of AFM spintronics. In this work, we study theoretically the Néel-type domain wall pinning and depinning at a notch in an antiferromagnetic (AFM) nano-ribbon. The depinning field depending on the notch dimension and intrinsic physical parameters are deduced and also numerically calculated. Contrary to conventional conception, it is revealed that the depinning field is remarkably dependent of the damping constant and the time-dependent oscillation of the domain wall position in the weakly damping regime benefits to the wall depinning, resulting in a gradual increase of the depinning field up to a saturation value with increasing damping constant. A one-dimensional model accounting of the internal dynamics of domain wall is used to explain perfectly the simulated results. It is demonstrated that the depinning mechanism of an AFM domain wall differs from ferromagnetic domain wall by exhibiting a depinning speed typically three orders of magnitude faster than the latter, suggesting the ultrafast dynamics of an AFM system.

Email: qinmh@scnu.edu.cn

Antiferromagnetic (AFM) materials are promising for next generation of spintronic devices and attract substantial attention especially because they have strong anti-interference capability and promised ultrafast magnetic dynamics.[1-8] As a frontier and highly concerned issue for advanced spintronics, the domain wall (DW) dynamics of antiferromagnets is under extensive investigation. Specifically, several stimuli have been proposed to drive the domain wall motion, including the Néel spin-orbit torques,[9-10] spin waves,[11-12] temperature gradients[13-15] and so on.[16-18] These works provide useful information for future AFM storage device design.

Nevertheless, most of these works discuss models on perfect samples and the wall pinning caused by disorder and local defects is neglected. As a matter of fact, the wall pinning may play an important role in magnetic dynamics. On one hand, for a realistic spintronic device where inhomogeneity and lattice defects are inevitable, the wall dynamics could be significantly affected and the wall pinning/depinning becomes the limited step for device operation. For example, it was reported that electrical current induced switching of AFM domains in CuMnAs occurs only in localized regions, strongly suggesting the important role of wall pinning.[19] Given these reasons, a clarification of the underlying mechanisms for wall pinning/depinning becomes essential. On the other hand, artificial lattice defects such as notches with proper shape could be used in discretizing domain wall position and enhancing its stability against thermal fluctuations and stray fields in potential race-track memory and logic devices.[20-24] Therefore, the dynamics of AFM domain wall pinning/depinning appears to be one of the core issues for application potentials and basic research of AFM spintronics.

Fortunately, the domain wall pinning in ferromagnetic systems have been extensively investigated, and the accumulated experience can be partially transferred to the study of AFM domain dynamics.[25-32] For a ferromagnetic domain wall, the depinning field can be analytically obtained by minimizing the total energy, demonstrating the critical role of notch geometry in pinning the wall.[26] More interestingly, the dependence of depinning field on the Gilbert damping for a ferromagnetic system has been revealed in micromagnetic simulations, and the damping constant, if small, can reduce the depinning field, contrary to the general expectation that they should be independent of each other.[27] This phenomenon not only reveals the complexity of domain wall pinning, but more importantly provides a method of domain wall manipulation. However, as far as we know, few work on the pinning/depinning of an AFM domain wall has been available, while this issue is certainly more important than and distinctly different from the case of ferromagnetic wall.

In proceeding, we may discuss the domain wall pinning/depinning for an AFM nanostructure with a notch, without losing the generality, while the calculation methods and main conclusions apply to antiferromagnets with other lattice defects. For simplicity consideration, such a notch has a rectangular section, as shown in Fig. 1(a). We can derive the depinning field $h_{dep}$ as a function of the notch size and uniaxial anisotropy in a simplified framework and the theory agrees well with numerical simulations in large damping systems. Moreover, it will be shown that the depinning field gradually increases to a saturation value with increasing damping constant, and this prediction allows one to modulate the damping constant through elaborately material design, so that the domain wall depinning can be in turn effectively controlled. In order to understand the underlying physics better, we perform the analytical calculation based on the one-dimensional model which reveals the important role of the internal domain wall dynamics. Our work also proposes a depinning mechanism for an AFM wall different from ferromagnetic wall. This new mechanism allows the depinning speed to be typically three orders of magnitude faster than that for a ferromagnetic wall depinning.

We start from the domain wall pinning at a rectangular notch for an AFM nanoribbon. This nanoribbon is geometrically defined by length $l$ along the $z$-axis, width $w$, and thickness $t_l$, as shown in Fig. 1. We discuss the scenario of current induced Néel spin-orbit torques (or staggered effective field), as demonstrated in CuMnAs and Mn$_2$Au for driving the domain wall motion, i.e. the wall is typically of the Néel type.[6,8-9] For this scenario, the model Hamiltonian is given by[33-34]

$$H = \frac{A_0}{2}\mathbf{m}^2 + \frac{A}{2}\nabla\mathbf{n}\cdot\nabla\mathbf{n} + L_0\mathbf{m}\cdot\nabla\mathbf{n} - \frac{K_z}{2}n_z^2 + \gamma\rho h n_z, \qquad (1)$$

where $A_0 = 4JS^2/a$ is the homogeneous exchange constant with AFM coupling $J > 0$, spin length $S$ and lattice constant $a$, $\mathbf{m}$ is the total magnetization $\mathbf{m} = (\mathbf{m}_1 + \mathbf{m}_2)/2S$ with $\mathbf{m}_1$ and $\mathbf{m}_2$ the AFM sublattice magnetizations, $A = 2aJS^2$ is the inhomogeneous exchange constant, $\mathbf{n}$ is the staggered magnetization $\mathbf{n} = (\mathbf{m}_1 - \mathbf{m}_2)/2S$, $L_0 = 2JS^2$ is the parity-breaking parameter, $K_z$

$= 2K_0S^2/a$ is the anisotropy constant along the z-axis in the continuum model with anisotropy constant $K_0$ in the discrete model, $\gamma$ is the gyromagnetic ratio, $\rho = S\hbar/a$ is the density of the staggered spin angular momentum per unit cell, $h$ is the staggered effective field and $n_z$ is the z component of **n**. Here, the notch has its width $d$ and depth $w_N$, as depicted in Fig. 1(a).

Noting that **m** is just a slave variable of **n**,[33] and we eliminate **m** by $\mathbf{m} = -L_0\nabla\mathbf{n}/A_0$ and obtain

$$H = \frac{A^*}{2}\nabla\mathbf{n}\cdot\nabla\mathbf{n} - \frac{K_z}{2}n_z^2 + \gamma\rho h n_z, \tag{2}$$

where $A^* = A - L_0^2/A_0$ is the effective exchange constant. As shown in the Supplementary Materials for the detailed derivation, the depinning field $h_{dep}$, based on this Hamiltonian model, can be solved strictly after a similar derivation[26]

$$h_{dep} = \frac{2K_0/\mu_S}{2w/w_N - 1}, \tag{3}$$

where $\mu_S$ is the saturation moment. It's noted that for an ultra-thin nanoribbon, the depinning field is independent of thickness. As clearly indicated in Eq. (3), $h_{dep}$ depends on several parameters including the anisotropy constant $K_0$ and the $w/w_N$ ratio. Thus, the devices with various depinning fields could be designed through modulating ratio $w/w_N$ and/or choosing appropriate materials.

In order to check the validity of Eq. (3), we also perform the numerical simulations based on the atomistic Landau-Lifshitz-Gilbert (LLG) equation,[14]

$$\frac{\partial \mathbf{S}_i}{\partial t} = -\frac{\gamma}{(1+\alpha^2)}\mathbf{S}_i \times [\mathbf{H}_i + \alpha(\mathbf{S}_i \times \mathbf{H}_i)], \tag{4}$$

where $\mathbf{S}_i$ is the normalized atomic spin at site $i$, $\alpha$ is the damping constant, $\mathbf{H}_i = -\mu_S^{-1}\partial H/\partial \mathbf{S}_i$ is the effective field. Without loss of generality, $l = 120a$, $t_l = a$, $w = 8a$, $K_0 = 0.02J$, $d = 4a$, $w_N = 2a$ and $\alpha = 0.02$ are selected, as shown in Fig. 1.

Fig. 1 presents the spin structures of the nanoribbon for various $h$. Here, the Néel-type AFM domain wall is clearly pinned at the notch at $h = 0$ and the spin configuration is symmetric around the notch due to the absence of chirality, as shown in Fig. 1(a). The spins on the wall mid-plane are aligned in parallel to the x-axis and perpendicular to those spins inside the AFM domains aside.

When a small $h$ is applied along the $z$-axis, the wall slightly shifts toward the right side, as seen from the delicate change of the spin configuration. With increasing $h$, those spins on the left side of the notch mid-plane tend to rotate towards the negative $z$-axis while those on the right side of the notch mid-plane tend to rotate towards the $x$-axis, as shown in Fig. 1(b) and 1(c), a consequence of the wall depinning from the notch. The wall depinning becomes clear in Fig. 1(c) where the wall mid-plane deviates clearly from the notch mid-plane. The spin configuration after the full wall depinning from the notch is shown in Fig. 1(d).

Subsequently, we investigate the dependences of $h_{dep}$ on the notch geometry and several physical parameters including the anisotropy and damping constants. The calculated curves (analytical) from Eq. (3) plus the simulated results (numerical) based on the LLG dynamics, Eq. (4), for different values of notch depth $w_N$, nanoribbon thickness $w$, anisotropy constant $K_0$, and damping constant ($\alpha$) are plotted in Fig. 2(a) ~ (d) respectively. Several features deserve highlighting here. First, the model calculated curves and numerically simulated data on dependences $h_{dep}(w_N)$, $h_{dep}(w)$, and $h_{dep}(K_0)$ respectively show qualitatively similar tendencies, suggesting that Eq. (3) can describe roughly these dependences although quantitative difference between the model and simulation appears for each dependence. Second, qualitative difference between the model and simulation appears for function $h_{dep}(\alpha)$, as shown in Fig. 2(d). While the model suggests independence of $h_{dep}$ on damping constant $\alpha$, the numerical simulation reveals that $h_{dep}$ is remarkably dependent of $\alpha$ in the small $\alpha$ regime. $h_{dep}$ shows a gradual growth with $\alpha$ until the large $\alpha$ regime where $h_{dep}$ becomes saturated, i.e. independent of $\alpha$ in the large $\alpha$ regime. The difference between Eq. (3) and simulated results for $h_{dep}(\alpha)$ is understandable since the LLG damping is a time-dependent effect. It is noted that the internal dynamics of domain wall is completely neglected in deriving Eq. (3), while this dynamics becomes particularly remarkable in the small $\alpha$ regime where the time-dependent spin oscillation can be significant due to the weak damping. Therefore, the model prediction Eq. (3) becomes invalid and the underlying physics should be reconsidered.

In order to uncover the intriguing physics, we need to track the domain wall evolution. In proceeding, we first define the position of a domain wall. Similar to the well-studied skyrmions, the position of a domain wall is estimated by $q(t)$[35]

$$q = \frac{\int z(1-|n_z|)dxdz}{\int (1-|n_z|)dxdz},\tag{5}$$

where $q$ is the coordinate of the wall mid-plane. Given this definition, one starts with the one-dimensional model with inclusion of the internal dynamics of domain wall motion.[28-29] The Hamiltonian density for this model reads[33]

$$H_{1D} = \frac{A_0}{2}\mathbf{m}^2 + \frac{A}{2}(\partial_z \mathbf{n})^2 + L_0 \mathbf{m}\cdot\partial_z \mathbf{n} - \frac{K_z}{2}n_z^2 + \gamma\rho h n_z + V(z),\tag{6}$$

where the pinning effect from the notch is described by potential energy $V(z)$.

Subsequently, we study the Lagrangian density $L = K - H_{1D}$ with $K = \rho\,\mathbf{m}\cdot(\dot{\mathbf{n}}\times\mathbf{n})$ is the kinetic energy term introduced by the Berry phase, and $\dot{\mathbf{n}}$ represents the derivative with respect to time.[33,36-37] Then, we eliminate $\mathbf{m}$ with $\mathbf{m} = (\rho\,\dot{\mathbf{n}}\times\mathbf{n} - L_0\partial_z\mathbf{n})/A_0$,[33] and obtain

$$L = \frac{\rho^2}{2A_0}\dot{\mathbf{n}}^2 - \frac{A^*}{2}(\partial_z\mathbf{n})^2 + \frac{K_z}{2}n_z^2 - \gamma\rho h n_z - V(z),\tag{7}$$

It is noted that the Rayleigh function density $R = \alpha\rho\,\dot{\mathbf{n}}^2/2$ is introduced into the Lagrangian formalism in order to describe the dissipative dynamics.[36-37] Following the earlier work, we assume a robust domain wall structure which can be described by $\mathbf{n} = [\text{sech}((z-q)/\lambda)\cos\Phi, \text{sech}((z-q)/\lambda)\sin\Phi, \tanh((z-q)/\lambda)]$,[36] where the azimuthal angle $\Phi$ of the wall is introduced as the collective coordinates. After substituting the domain wall ansatz and applying the Euler-Lagrange equation, we obtain the equation of motion for variables $q$ and $\Phi$,

$$\frac{\rho^2}{\lambda A_0}\ddot{q} + \frac{\alpha\rho}{\lambda}\dot{q} + \frac{d\varepsilon}{dq} - \gamma\rho h = 0,\tag{8}$$

and

$$\frac{\rho^2}{A_0}\ddot{\Phi} + \alpha\rho\dot{\Phi} = 0,\tag{9}$$

It is noted that the first term in Eq. (8) describes the wall inertia and other terms represent the forces exerted respectively by the damping $\alpha$, pinning potential $\varepsilon(q)$, and current-induced effective magnetic field $h$. By substituting the initial condition $\Phi(0) = d\Phi/dt|_{t=0} = 0$ into Eq. (9), one obtains $\Phi(t) = 0$, consistent with the fact that an AFM domain wall is confined in the easy plane due to the antiparallel arrangement of neighboring spins.

For simplicity, we assume a parabolic potential[23,29]

$$\varepsilon(q) = \begin{cases} K_N q^2/2 & (|q| < L_N) \\ K_N L_N^2/2 & (|q| \geq L_N) \end{cases}, \qquad (10)$$

where $K_N$ is the elastic constant and $L_N$ is the radius of the potential well. After substitutions and necessary simplification, the equation of motion for $q$ is updated to

$$\ddot{q} + G\dot{q} + \omega_N^2 q - h_N = 0, \qquad (11)$$

where $G = \alpha A_0/\rho$, $h_N = \gamma A_0 \lambda h/\rho$, and $\omega_N = (\lambda A_0 K_N/\rho^2)^{1/2}$ is the natural angular frequency of the free harmonic oscillator. Here, we can see the existence of domain wall oscillation if damping constant $\alpha$ is small. This oscillation is the major reason for the invalid prediction of the depinning field by Eq. (3).

Noting that Eq. (11) describes the damping oscillation of a domain wall, one has the solution for $\alpha < \alpha_c = 2\rho^2 a \omega_N/JA_0$ representing the under-damped oscillation:

$$q(t) = e^{-Gt}\left(C_1 \cos\omega_p t + C_2 \sin\omega_p t\right) + h_N/\omega_N^2, \qquad (12)$$

where $\omega_p = (\omega_N^2 - G^2/4)^{1/2}$ is the oscillating angular frequency of the wall, and $C_1$, $C_2$ are integral constants depending on the initial condition.

For better illustration, the simulated $q(t)$ curves based on the LLG equation at various damping constant $\alpha$ are plotted in Fig. 3(a), benefiting to discussion. For $\alpha > 0.005$, one observes the domain wall oscillation around the equilibrium position with an attenuating amplitude. Moreover, the oscillation amplitude is enhanced with the decreasing $\alpha$. Finally, for $\alpha < 0.005$, when the maximum displacement of the wall oscillation, defined as $|\Delta q|_{max} = |q(t) - q(0)|_{max}$, exceeds the height of the pinning potential,[29] the wall would successfully depin from the notch and propagates freely along the nanoribbon.

As demonstrated in Eq. (12), the displacement of the wall oscillation consists of the oscillatory part ($A_S$) and stationary part ($q_{eq}$),[29] and its maximum value is approximately given by

$$|\Delta q|_{max} = A_S + q_{eq} = e^{-G\arctan(C_2/C_1)/\omega_p}\sqrt{C_1^2 + C_2^2} + h_N/\omega_N^2, \qquad (13)$$

where $\omega_p \approx \omega_N$ is obtained for $\alpha < \alpha_c$. In this case, since $|\Delta q|_{max}$ decreases exponentially with $\alpha$, larger external field is required to generate the wall displacement for the wall depinning. As $|\Delta q|_{max} > L_N$, the wall eventually depins from the notch.

Noting that the pinning potential parameters including $K_N$ and $L_N$ are unknown, we need a reasonable estimation of them by fitting the simulated results based on Eq. (13). As shown in Fig. 3(b) where the simulated furthest position of the domain wall, $q_{max}$, as a function of $\alpha$, is plotted. The excellent fitting of the simulated data by Eq. (13) on the other hand further confirms the validity of our theory.

Since the oscillating amplitudes $C_1$ and $C_2$ are proportional to external or current induced field $h$, one can introduce the field-independent parameters $c_1 = C_2/C_1$, $c_2 = (C_1^2 + C_2^2)^{1/2}/h$ for brevity. Subsequently, the depinning field under the condition $|\Delta q|_{max} = L_N$ is obtained:

$$h_{dep} = \frac{L_N}{e^{-G \arctan c_1 / \omega_p} c_2 + \gamma \rho / K_N}, \qquad (14)$$

Similar fitting approach can be used to estimate $L_N$. As shown in Fig. 2(d), the simulated results coincide very well with Eq. (14) with one adjustable variable $L_N$, demonstrating the important role of the domain wall oscillation in the domain wall depinning. Such an oscillation behavior is one character of the internal dynamics for a AFM nanoribbon with a notch.

Finally, we would like to address the significance of the present results. It is known that the performance of domain wall based race-track memory not only depends on the wall motion velocity, but also relies on the wall depinning time. It is clearly shown here that an AFM domain wall depinning is distinctly different from that of a ferromagnetic domain wall. For a ferromagnetic nanoribbon, the wall oscillation is related to the wall internal angle which is mainly determined by the internal fields including magnetocrystalline anisotropy and Dzyaloshinskii-Moriya (DM) exchange.[27] Generally, the depinning time is inversely proportional to the magnitude of internal fields and has a typical value of ~ 1.0 ns.[22-24,27] However, for an AFM system, the wall oscillation stems from the second-order derivative of DW position $q$ with respect to time rather than the azimuthal angle of the DW, as clearly illustrated in Eq. (8). Since the derivative originates from the strong AFM exchange interaction between two sublattices which is about three orders larger than the anisotropy and DM exchange, one is sure that the depinning time for such an AFM domain wall should be three orders of magnitude shorter than a ferromagnetic one. It implies a surprisingly short depinning time of ~ 0.001 ns for CuMnAs with the Néel temperature $T_N \approx 480$ K, $a \approx 3.8$ Å and $\mu_S \approx 3.6$ $\mu_B$,[38] where $\mu_B$ is the Bohr magneton. While it is well believed that the AFM domain switching is faster than ferromagnetic domain switching, the present work presents a quantitative estimation of the domain wall depinning time, direct evidence with this well-believed but not yet well-evidenced claim.

In conclusion, we study theoretically the domain wall pinning and depinning at a notch in an AFM nano-ribbon. The depinning field depending on the notch dimension and intrinsic physical parameters are derived theoretically and also simulated based on the LLG equation. Contrary to the conventional conception, the remarkable dependence of the depinning field on the damping constant is revealed, which attributes to the time-dependent oscillation of the DW position in the small damping region. A one-dimensional model considering the internal dynamics of DW is investigated theoretically to explain perfectly the simulations. More importantly, our work also demonstrates the different depinning mechanism of an AFM DW from FM DW which may result in a depinning speed typically three orders faster than the latter, demonstrating again the ultrafast dynamics of an AFM system.

**Acknowledgment**

We sincerely appreciate the insightful discussions with Zhengren Yan, Yilin Zhang and Huaiyang Yuan. The work is supported by the National Key Projects for Basic Research of China (Grant No. 2015CB921202), and the Natural Science Foundation of China (No. 11204091), and the Science and Technology Planning Project of Guangdong Province (Grant No. 2015B090927006), and the Natural Science Foundation of Guangdong Province (Grant No. 2016A030308019).

**FIGURE CAPTIONS**

Fig.1. (color online) Equilibrium spin structures around the notch in the AFM nanoribbon with lattice sizes $l \times w \times t_l$ under (a) $h = 0$, (b) $h = 0.002J/\mu_S$, (c) $h = 0.004J/\mu_S$, and (d) $h = 0.00458J/\mu_S$. The color represents the magnitude of the $z$ component of the staggered magnetization $n_z$, and the position of the DW center is depicted by the black dashed lines.

Fig.2. (color online) Numerical (empty circles) and analytical (blue solid line) calculated depinning field as a function of (a) the depth of the notch $w_N$, (b) the width of the nanoribbon $w$, (c) the anisotropy constant $K_0$, and (d) the damping constant $\alpha$. The red solid line in (d) is the fitting results based on Eq. (14).

Fig.3. (color online) (a) The DW position as a function of time for various damping constants under $h = 0.0039J/\mu_S$. (b) Numerical (empty circles) and analytical (solid line) calculated maximum displacement of the DW as a function of $\alpha$ under $h = 0.0039J/\mu_S$.

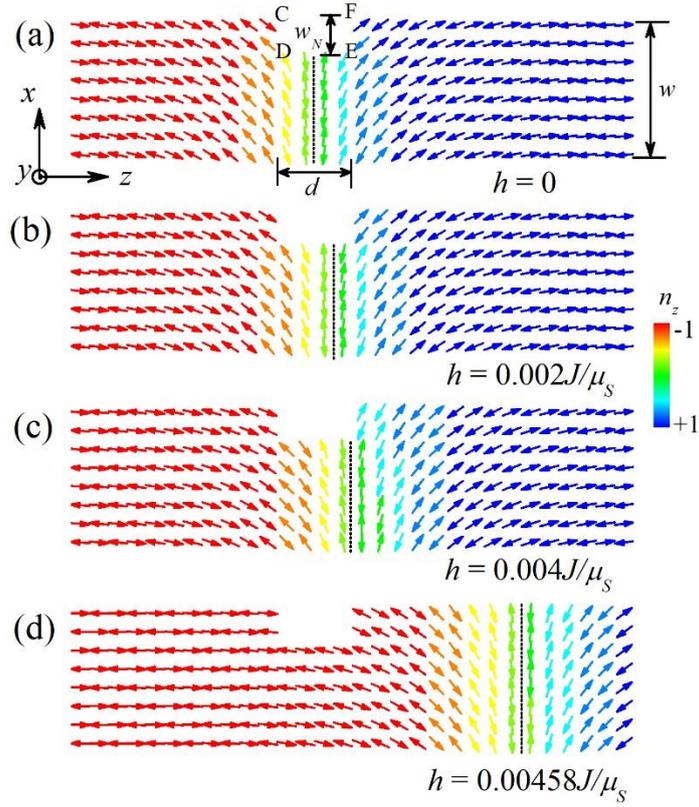

Fig.1. (color online) Equilibrium spin structures around the notch in the AFM nanoribbon with lattice sizes $l \times w \times t_l$ under (a) $h = 0$, (b) $h = 0.002J/\mu_S$, (c) $h = 0.004J/\mu_S$, and (d) $h = 0.00458J/\mu_S$. The color represents the magnitude of the $z$ component of the staggered magnetization $n_z$, and the position of the DW center is depicted by the black dashed lines.

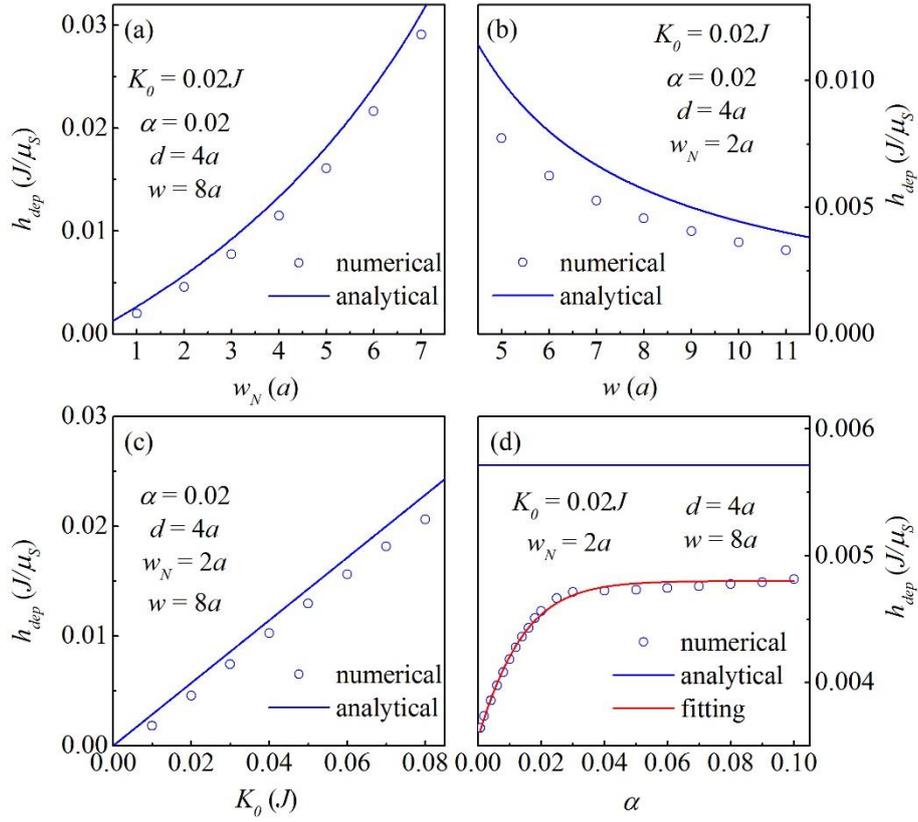

Fig.2. (color online) Numerical (empty circles) and analytical (blue solid line) calculated depinning field as a function of (a) the depth of the notch $w_N$, (b) the width of the nanoribbon $w$, (c) the anisotropy constant $K_0$, and (d) the damping constant $\alpha$. The red solid line in (d) is the fitting results based on Eq. (14).

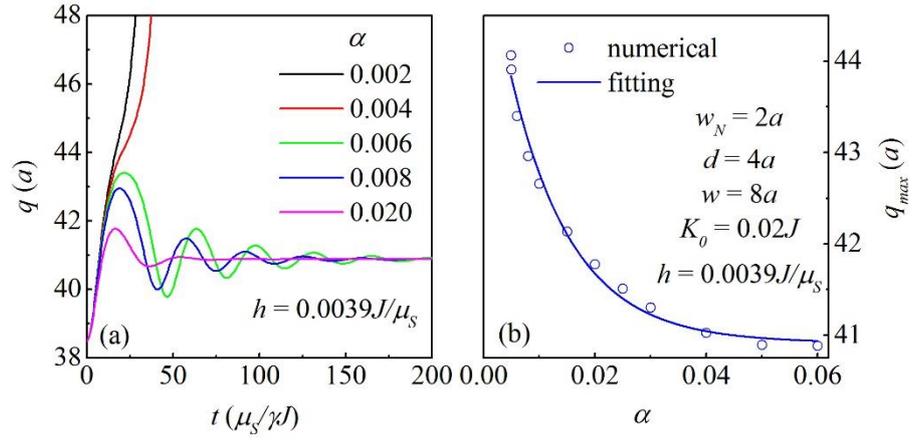

Fig.3. (color online) (a) The DW position as a function of time for various damping constants under $h = 0.0039J/\mu_S$. (b) Numerical (empty circles) and analytical (solid line) calculated maximum displacement of the DW as a function of $\alpha$ under $h = 0.0039J/\mu_S$.

# Supplementary material for

# "Depinning of domain walls in notched antiferromagnetic nanostructures"

Z. Y. Chen[1], M. H. Qin[1,*], and J. –M. Liu[2]

[1]*Institute for Advanced Materials, South China Academy of Advanced Optoelectronics and Guangdong Provincial Key Laboratory of Quantum Engineering and Quantum Materials, South China Normal University, Guangzhou 510006, China*

[2]*Laboratory of Solid State Microstructures and Innovative Center for Advanced Microstructures, Nanjing University, Nanjing 210093, China*

## A. Derivation of the depinning field

The model Hamiltonian density reads

$$H = \frac{A_0}{2}\mathbf{m}^2 + \frac{A}{2}\nabla\mathbf{n}\cdot\nabla\mathbf{n} + L_0\mathbf{m}\cdot\nabla\mathbf{n} - \frac{K_z}{2}n_z^2 + hn_z. \tag{1}$$

After eliminating $\mathbf{m}$ with $\mathbf{m} = -L_0\nabla\mathbf{n}/A_0$, we obtain

$$H = \frac{A^*}{2}\nabla\mathbf{n}\cdot\nabla\mathbf{n} - \frac{K_z}{2}n_z^2 + \gamma\rho h n_z. \tag{2}$$

In the following, we use the same method with Ref. 24 to derive the depinning field for AFM DWs. At low temperatures, we introduce the Lagrange multiplier $\xi$ to take into account the constraint condition $\mathbf{n}\cdot\mathbf{n} = 1$, and then construct a new function

$$F = \int dV\left(\frac{A^*}{2}\nabla\mathbf{n}\cdot\nabla\mathbf{n} + f_{ot}\right) - \xi(\mathbf{n}\cdot\mathbf{n} - 1), \tag{3}$$

where $f_{ot}$ is the sum of the anisotropy and Zeeman energy. Using the variational method, we obtain

$$-l_{ex}^2\nabla^2 n_i + \frac{\partial f_{ot}}{\partial n_i} + 2\xi n_i = 0, \tag{4}$$

where $l_{ex} = (aA^*/J)^{1/2}$ is the exchange length in AFM systems, $n_i$ is the $x_i$ component of $\mathbf{n}$ ($x_i$

---

[*]qinmh@scnu.edu.cn

$= x, y, z$). To eliminate $\xi$, we take the product of Eq. 4 and sum over $i$ and obtain

$$-l_{ex}^2 \frac{\partial n_i}{\partial x_j} \nabla^2 n_i + \frac{\partial f_{ot}}{\partial x_j} = 0. \tag{5}$$

Transforming Eq. 5 with the identity

$$\frac{\partial g}{\partial x_j} \nabla^2 g = \nabla \cdot \left( \frac{\partial g}{\partial x_j} \nabla g \right) - \frac{1}{2} \frac{\partial}{\partial x_j} (\nabla g)^2, \tag{6}$$

and we obtain

$$\frac{\partial}{\partial x_j} \left( \frac{1}{2} l_{ex}^2 \nabla n_i \cdot \nabla n_i + f_{ot} \right) = l_{ex}^2 \nabla \cdot \left( \frac{\partial n_i}{\partial x_j} \nabla n_i \right). \tag{7}$$

To eliminate the space-dependent variables, we take the summation over the whole regions of the sample $\Omega$,

$$\int_\Omega dV \frac{\partial}{\partial x_j} \left( \frac{1}{2} l_{ex}^2 \nabla n_i \cdot \nabla n_i + f_{ot} \right) = \int_\Omega dV l_{ex}^2 \nabla \cdot \left( \frac{\partial n_i}{\partial x_j} \nabla n_i \right) = \int_{\partial\Omega} d\mathbf{S} \cdot l_{ex}^2 \frac{\partial n_i}{\partial x_j} \nabla n_i, \tag{8}$$

where $\partial\Omega$ is the boundary of $\Omega$. Considering the boundary condition $\nabla n_i = 0$, we have

$$\int_\Omega dV \frac{\partial}{\partial z} \left( \frac{1}{2} l_{ex}^2 \nabla n_i \cdot \nabla n_i + f_{ot} \right) = \int_{GH-EF+CD-AB} dxdy \left( \frac{1}{2} l_{ex}^2 \nabla n_i \cdot \nabla n_i + f_{ot} \right) = 0. \tag{9}$$

Substituting the configuration of the system into Eq. 9 and we obtain

$$t_l \int_{CD-EF} dy \left( \frac{1}{2} l_{ex}^2 \nabla n_i \cdot \nabla n_i + f_{ot} \right) = -2ht_l w. \tag{10}$$

Then the magnitude of the current-induced effective field is given by

$$h = \frac{t_l \int_{CD-EF} dy \left( \frac{1}{2} l_{ex}^2 \nabla n_i \cdot \nabla n_i + f_{an} \right)}{t_l w_N \left( 2w/w_N + \langle n_{zEF} \rangle - \langle n_{zCD} \rangle \right)}, \tag{11}$$

where $\langle n_{zEF} \rangle$, $\langle n_{zCD} \rangle$ are the average $z$ components of $\mathbf{n}$ on surfaces $EF$ and $CD$, respectively.

The depinning field represents the minimum field to move a DW, and in other words, the maximum field that Eq. 11 has a stationary solution. Thus, critical condition is the key to deriving the depinning field. Similar to the earlier work, we consider the critical condition $\langle n_{zEF} \rangle = 0$, $\langle n_{zCD} \rangle = 1$ in our derivation, whose validity is confirmed in Fig. 1(c) in the manuscript. After substitutions and simplifications, we obtain the depinning field of AFM

DWs

$$h_{dep} = \frac{2d_z / \mu_S}{2w / w_N - 1}. \tag{12}$$